\documentclass[8pt,a4paper]{scrartcl}
\usepackage[top=2.5cm,bottom=2.5cm,left=2cm,right=2cm]{geometry}

\usepackage{rotating}
\usepackage{amsmath}
\usepackage{color}
\usepackage{graphicx}
\usepackage[square, numbers]{natbib}
\usepackage{caption}
\usepackage{nonfloat}

\usepackage{cancel}
\usepackage{mathcomp}
\usepackage{braket}
\usepackage{multicol}
\usepackage[applemac]{inputenc}
\usepackage{color}
\usepackage{shadow}
\usepackage{times}
\usepackage[superscript,nospace]{cite}

\usepackage{booktabs}
\usepackage{fancyhdr}
\newcommand\myfigure[1]{%
\medskip\noindent\begin{minipage}{\columnwidth}
\centering%
#1%
\end{minipage}\medskip}

\newcommand\myfigurelarge[1]{%
\end{multicols}
\medskip\noindent\begin{minipage}{\columnwidth}
\centering%
#1%
\end{minipage}\medskip\begin{multicols}{2}}

\newcommand{\be}{\begin{equation}}
\newcommand{\ee}{\end{equation}}
\newcommand{\bea}{\begin{eqnarray}}
\newcommand{\eea}{\end{eqnarray}}

\begin{document}





\newcommand{\grenoble}{Institut N\'eel,
CNRS/UJF, 25 rue des Martyrs BP 166, B\^{a}timent D 38042 Grenoble
cedex 9 France} 
\newcommand{\nice}{Universit\'e de Nice - CNRS UMR 7335,  Institut Non Lin\'eaire de Nice, 1361
route des lucioles, 06560 Valbonne, France}
\begin{center}
~ \vspace{0.7cm}\\
{\Large{\bf
Trends in condensed matter physics: is research going faster and faster?}} \\     
\medskip
{\bf{\underline{C. Attaccalite} \footnote{\grenoble}}, S. Barland\footnote{\nice} }\\ 
{\emph{\grenoble \\ \nice}}\medskip\\         
{Received 00.00.2010, accepted 00. 00. 2010, published 00.00.2010}
 \bigskip
\end{center}


\begin{center}
\begin{minipage}[c]{0.7\textwidth}
In this paper we study research trends in condensed matter physics. Trends are analyzed by means of the the number of publications in the different sub-fields as function of the years. We found that many research topics have a similar behavior with an initial fast growth and a next slower exponential decay. We derived a simple model to describe this behavior and built up some predictions for future trends. 

\end{minipage}
\end{center}
\bigskip

\begin{multicols}{2}

\section{Introduction}
The investigation of social phenomena has acquired a large importance in the last years.\cite{RevModPhys.81.591} In particular statistical physics has been applied to study interaction of individuals as elementary units of a social structure.
In these models each individual interacts with a limited number of peers usually negligible compared with the total number of agents.  In this ensemble of interacting agents, the statistical physics has been used to investigate opinion and cultural dynamics, crowds behavior, social spreading phenomena and so on.\cite{RevModPhys.81.591}\\
On the other hand the structure of the human connections has been subject of an intense study by means of networks theory.\cite{RevModPhys.74.47}  It has been shown that humans form particular kinds of networks, the so-called scale-free networks, that exhibit a very short path from one site to another thanks to the presence of large hubs connected to many nodes.\cite{RevModPhys.81.591} Also scientists and the scientific community have been subject of similar studies.
In particular network theory has been employed to show that scientific collaborations\cite{newman} form a scale-free network. One important characteristic of these networks is that new ideas and trends can spread without any initial threshold  due to the unbounded  fluctuations in the connectivity.\cite{PhysRevLett.86.3200} \\ 
In this paper we will analyze the birth and death of trends in the scientific community, and in particular in condensed matter physics. In the last fifty years different topics have been subject of intense research in condensed matter physics, ranging from superconductivity, to charge density waves or nanostructures. Researchers, during  their career,  have usually worked on some of these subjects moving  from one to another when they found the new one more stimulating or challenging. This collective behavior gave rise to the birth, growth and death of new trends.  Here we want to investigate this dynamics without  dealing directly with the diffusion of ideas in the scientists network, but devising a simple model to describe birth and death of new research topics in the community. In order to do so we will perform a bibliometric study based on lexical analysis of papers published in the last fifty years. Bibliometrics has been already employed to detect and investigate emerging research topics\cite{glanzel2012bibliometric} by mean of co-citation links or text-based techniques.\cite{springerlink:10.1007/BF02017232} 
Here we will use a simple text-based approach to obtained the data that will then analyzed by mean of a model derived by population dynamics. The model resemblances the dynamics of fashion in social science, that has been studied by means of Langevin  and Fokker-Planck equations.\cite{weidlich,Nakayama} We will show that different sub-fields, irrespective of their importance and time location, exhibit a similar behavior that is well described by our simple model. \\
The questions we would like to answer are: Are researches changing topics more often than in the past? Do research trends last less then in the past? And if this is true, is it due to the larger number of researchers? We do not have a definite answer to all these questions but we will show that with a simple  model we can get some insights for trends in condensed matter physics.
The paper is organized as follows: in section \ref{data} we discuss publications data; in section \ref{modelsection} we present the model used to describe trends in condensed matter physics; in section \ref{fit} we apply our model to the publications data and report the  results; finally in section \ref{conclusion} we draw conclusions and discuss open questions.

\section{Publications and data}
\label{data}
The number of scientific publications in physics is exponentially growing (see Fig.~\ref{normfigure}). This trend has been found also in other research fields as chemistry, biology and so on.\cite{expgrowth} However the growth pattern of given sub-fields may be different from the total one, as already pointed out in previous studies~\cite{expgrowth}. This behavior is due to the fact that particular research areas are saturated, lack of interest or possible innovation.\\
In order to evaluate trends in condensed matter physics we consider the ratio between the number of publications in a given sub-field and the total one.
The ratio is a good indicator for the importance of a given research sub-field because it is independent from the total publications number and it allows to compare research topics at different times.
\myfigure{
    \vspace{5 pt}
    \includegraphics[width=0.6\textwidth,angle=-90]{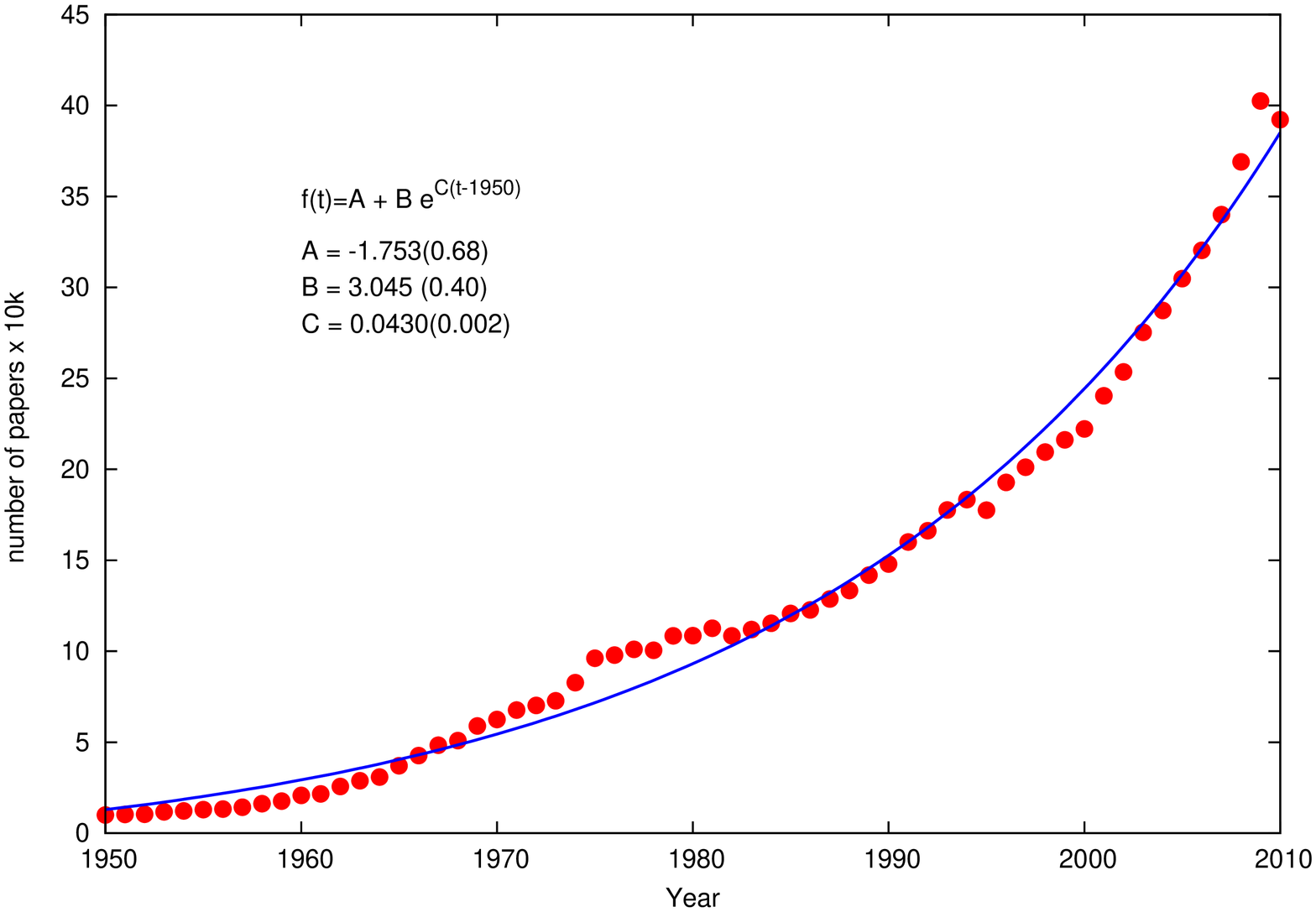}
\figcaption{(color online) Normalization factor obtained from SAO/NASA Astrophysics Data System Abstract Service\cite{SAO}. The total number of papers has been fitted with an exponential curve. Fit parameters are reported in the figure.}
\label{normfigure}}
We started analyzing the number of abstracts  published in physics from 1950 to 2011.\cite{data}  We found an exponential growth with a double time of about 16 years (data are shown in Figure \ref{normfigure}) in agreement with previous studies.\cite{0034-4885-32-2-306,expgrowth}  Then we extracted the number of papers in the different sub-fields, and normalized this number with the total number of papers published each year. Some results are reported in Figure~\ref{example4}.
The selection of papers belonging to a given sub-field has been done by means of a search in the abstract database using combination of several keywords to best represent a research topic (see for instance the first column of Table~\ref{table1}). 

\section{The model}
\label{modelsection}
In order to analyze trends in condensed matter physics we use a modified version of the logistic equation.\cite{logistic} In his original formulation Pierre-Fran\c{c}ois Verhulst applied the logistic equation to the population growth, assuming the reproduction rate being proportional to both the existing population and the amount of available resources:
\be
\frac{dP}{dt} = r P \left ( 1 - \frac{P}{K} \right),
\label{logistic}
\ee
where $r$ is the growth rate and the $K$ represents the environment carrying capacity.\\
In our model we consider $P$ as the fraction of scientists that decides to work on a given research topic and we suppose $P$ to be proportional to the paper ratio. As in the logistic model we suppose that the number of scientists working on a sub-field increases due to the available resources. In our case the "resources" represent the possibility to make new discoveries in the sub-field, but also the possibility to get more grants and more citations in an emerging research area.
This gain is equivalent to the so-called "bandwagon" effect in fashion dynamics.\cite{Nakayama}\\
\myfigure{\includegraphics[width=1\textwidth]{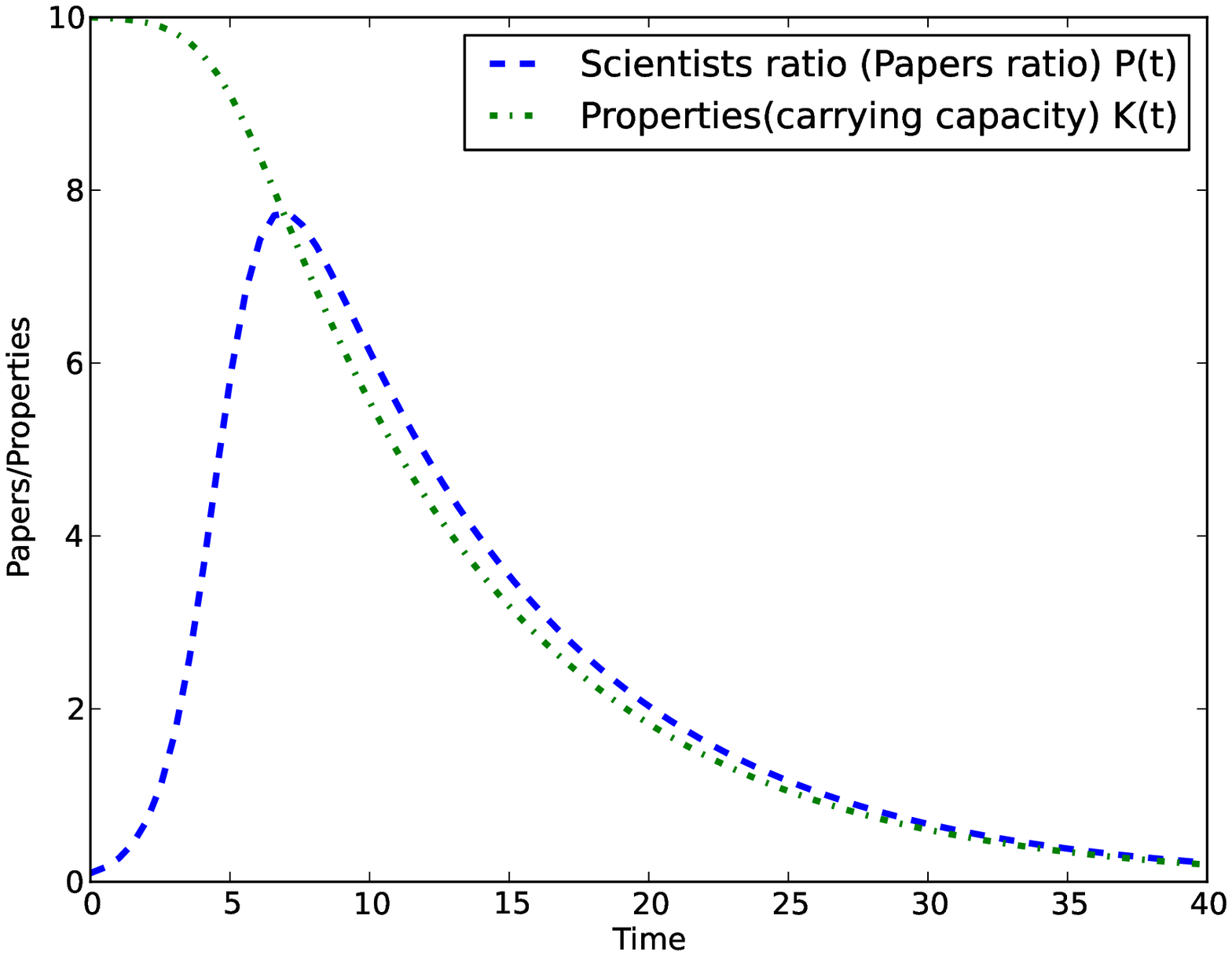}
\figcaption{(color online) Simple model for fashion in physics, with parameters $K(t=0)=10$, $P(t=0)=0.1$ and parameters $r=0.1$, $s=0.01$.}
\label{model}}

Differently from the logistic model we suppose that the carrying capacity of a given research topic decreases with time. In fact one can expect that after some time a large part of the possible experiments and theory will be already over. Moreover there is also an economical component\cite{economyscience} that contributes to the carrying capacity decrease: grants on a new and hot topic have clearly more successful probabilities than on an old and widely studied subject. This fact automatically decreases the possibility to have many scientists working on old research topics due to the reduction of available positions. All these effects, that contribute to reduce the carrying capacity are equivalent to the so-called "snob effect" in the fashion dynamics.\\
Therefore we suppose the $K$, the carrying capacity of our system, decreases proportionally to the number of scientists  working on a given research topic multiplied by a \emph{research rate coefficient} $s$
\be
\frac{dK(t)}{dt} = - s P(t).
\label{decrease}
\ee
The  \emph{research rate coefficient} $s$  represents the speed at which the properties $K(t)$ are investigated by scientists $P(t)$. In principle the decay of $K(t)$ can have a more complex dependence from $P(t)$. In literature there are several studies that include retardation effects or non-linear coupling,\cite{Yukalov20091752} however we did not find particular reasons for using more complicated models than a linear one. Combining equations~\ref{logistic} and \ref{decrease} we get:\\
\be
\frac{d^2K(t)}{dt^2} = r \frac{dK(t)}{dt}  \left ( 1+ \frac{1}{s} \frac{dK(t)}{dt} \frac{1}{K(t)}  \right).
\ee
The solution of this equation has the form:
\bea
P(t) &=& \frac{\left( -{\frac {rs}{ \left( r-s \right)  \left( {\it C_2}\,r+{\it C_1}\,{{\rm e}^{rt}} \right) }} \right) ^{{\frac {s}{r-s}}}r{\it C_1}\,{{\rm e}^{rt}}} {\left( r-s \right)  \left( {\it C_2}\,r+{\it C_1}\,{{\rm e}^{rt}} \right)}  \label{soleq1} \\
K \left( t \right) &=& \left( {\frac {rs}{ \left( -r+s \right)  \left( {\it C_2}\,r+{\it C_1}\,{{\rm e}^{rt}} \right) }} \right) ^{{\frac {s}{r-s}}} \label{soleq2}
\eea
where $C_1$ and $C_2$ are the integration constants that depend from $P(0)$ and $K(0)$. In order to understand the behavior of this model we consider the limits at short and long times of the previous solution.
For $t\rightarrow \infty$ we found that  eq.\ref{soleq1} decays exponentially as $P(t \rightarrow \infty) \simeq e^{-rs/(r-s) t}$. Since $s$ is supposed to be much smaller than $r$ the exponent reduces to  $P(t \rightarrow \infty) \simeq e^{-st}$. On the other hand for short time $t$, $P(t \rightarrow 0)$ goes as $P(t \rightarrow 0) \simeq c + e^{rt}$.  A typical sketch of the solution is reported in the Figure~\ref{model}.

From the limits at short and long time we can see that the initial growth is dominated by the $r$ coefficient while the exponential decay is dictated by the research rate coefficient $s$. In the language of population dynamics  the exponential decay is related to the decrease of carrying capacity, while the initial  growth is due to the  diffusion of a research topic in the scientific community. 
In order to analyze the publication data of different sub-fields we simplify eq.~\ref{soleq1} in such a way to catch short and long time behaviors with a simple formula:
\be
P_{model}(t)=P_0  \frac{1}{ e^{-\alpha (t - t_0) } + e^{\beta (t -t_0) }} + c.
\label{simplemodel}
\ee
\\
This model keeps only the main exponents for the growth and the decay part respectively, plus a constant $P_0$ that is the peak intensity while $t_0$ is related to the maximum position. We added a constant $c$ not present in the initial equation that will be discussed in the next section.\\
Notice that this model does not apply to any research topic, in fact there are some active fields that continue to grow without no sign of decay, as for instance research on Silicon, Density Functional Theory, Nanowires and so on (see Figure~\ref{example4}). Moreover this model can be easily generalized to take into account more peaks as for instance in superconductivity where the different discoveries gave rise to a sum of multiple curves, as the one in Figure~\ref{model}, translated in time and with different intensities. 

\myfigure{\small
    \begin{tabular}{ | c | c | c | c | c |}
    \hline
    Topic &  $ t_0 $ &  $(P_0/P_{sc})$ &   $\beta$ & $\alpha$ \\ \hline
    crystal fields  & $  1957.0(7) $ &$ 0.095(5) $ & $0.018(2)$ & $0.27(4)$\\
       helium-3/-4  & $ 1959(2)$ &$ 0.051(3) $ &$ 0.06(1)$ & $0.17(5)$ \\
    alkali halides  & $ 1960(1)$ &$ 0.186(6) $ &$ 0.070(6) $& $0.12(1)$ \\
        spin waves  & $ 1960.51(1)$ &$ 0.125(12) $ &$0.044(9) $ & $0.30(4)$\\
    Haas-van-Alphen  & $  1960.9(5) $ &$ 0.023(3) $ & $ 0.060(7) $ & $0.5(2)$\\
       Brullouin scattering & $ 1969(1)$ &$  0.048(3) $ &$ 0.024(4) $ & $0.24(3)$\\
         Compton scattering & $  1971.1(3)  $ &$ 0.0173(9)  $ &$ 0.060(4) $ & $0.8(1)$ \\
                spin glass  & $ 1977.9(3)$ &$     0.097(3) $ &$0.018(2) $ & $0.6(1)$\\
      charge density waves  & $1978.8(7)$ &$ 0.054(3) $ &$0.026(4)$ & $0.33(4)$ \\
        Josephson junctions & $ 1983(1)$ & $  0.11(1)$ &$ 0.038(8)$ & $0.15(1)$\\
        fract. Quantum Hall & $1983.8(3)$ & $ 0.0162(1) $ &$0.018(4)$ & $1.6(5)$\\
           four wave mixing & $  1984.5(5)$ & $ 0.091(5) $ &$  0.050(5)$ & $0.30(2)$\\
          quasicrystals     &$1985.7(2)$ & $ 0.031(2)$ &$0.025(4)$ & $1.6(5)$\\
              Penrose Tiles &$1986.1(3)$   &$ 0.022(1)$  &$0.023(5)$ & $1.0(2)$\\
          superconductivity &$1986.7(2)$   &$ 1.00(6)$  &$0.045(5)$ & $3(2)$\\
         slave bosons       &$1989.5(3)$   &$ 0.0133(4)$  &$ 0.062(6) $ & $0.66(7)$ \\
          t-J model         &$1990.1(1)$   &$ 0.037(1)$  &$ 0.090(5) $ & $1.2(1)$ \\
          fullerens         &$1991.2(1)$  &$ 0.169(4)$ &$0.030(2)$ & $2.8(4)$\\
           Luttinger liquid & $1993.1(3)$ & $ 0.024(1)$ &  $0.032(5)$ & $0.76(9)$\\
    giant-magnetores. & $1993.0(1)$ & $ 0.049(2)$ &  $0.071(7)$ & $3(1)$ \\
         Andreev reflection & $1993.0(6)$ & $  0.013(1)$ &  $0.014(8)$ & $0.54(9)$\\
         critical exponents & $1993(2)$ & $  0.163(7)$ &  $ 0.078(1) $ & $0.10(1)$\\
               Bethe ansatz & $1994(1)$ & $ 0.045(3)$ &  $0.08(1) $ & $0.19(1)$\\
         composite fermions & $1994.7(3)$ & $ 0.022(2)$ &  $0.11(1)$ & $1.4(3)$\\
         critical phenomena & $1995(2)$ & $ 0.096(4)$ &  $0.09(2)$ & $0.10(1)$\\
     spin-charge separation & $1999(1)$ & $ 0.0125(1)$ &  $  0.15(5)$ & $1.0(4)$ \\
          Coulomb blockade & $1999(1)$ & $ 0.069(6)$ &  $  0.08(1)$ & $0.26(2)$ \\
         nanotubes         &$2003.1(4)$   &$ 0.57(4)$  &$ 0.08(1)$ & $0.41(1)$\\
    \hline
    \end{tabular}
    \vspace{5 pt}
\tabcaption{
(color online) Parameters obtained by fitting the normalized number of abstracts with the model in Eq.~\ref{model}. The strength of each research topic has been normalized to the one of superconductivity $P_{sc} = 0.0262$. \label{table1}}
}

\section{Data fit and results}
\label{fit}
We suppose that the relevance of a research topic can be evaluated from the number of papers published on that subject. We extracted this number from the SAO/NASA Astrophysics Data System Abstract Service (SAS)\cite{SAO}. Since the total number of papers published every year is exponentially growing,  we normalized the number of papers of a given research area with the total one in order to evaluate the importance of each research topic. 
Also the normalization factor is obtained by the SAO/NASA database (see Fig.~\ref{normfigure}).
The data for the different topics were then fitted with the model described by Eq.~\ref{simplemodel}, where the constant $c$ has been always set to zero except for the case of "Superconductivity". In fact in this case there was already an active research community working on "BCS" superconductivity before the discovery of high-Tc superconductors boosted again the number of papers. In order not to miss papers and improve search results  we included a list of synonymous for each research area as for instance "superconductors, superconductor, superconductivity etc....".  The fit results are reported in Table~\ref{table1}, where we renormalized the peak intensity with the one of "superconductivity", the larger peak in our sample, in such a way to underline the relevance of the each sub-field. Selected data and relative fits are also plotted in Fig.~\ref{example4}.\\
Notice that recent research topics exhibit larger error bars because we could fit only the beginning of the decay tail. Moreover some active sub-fields as "graphene, topological insulators, quantum computing, photovoltaic, etc..."  are sill in the growing part of the curve or have just started to decrease and this makes impossible to perform an accurate fit with our model.

In Table \ref{table1} we report the different parameters entering in the model that describes trends in condensed matter physics. 
Unfortunately the results for the $\alpha$ parameters are very noisy due to the fast diffusion of new research topics in the scientific community.
This makes more difficult to get an accurate estimate for this exponent. Nevertheless we found that $\alpha$ is slowing increasing with time($t_0$), and this indicates a faster spread of ideas among scientists. This fast spread can be explained by the fact that the structure of the collaboration networks is scale-free\cite{newman} and in this kind of networks there is a vanishing threshold to spread ideas and trends.\cite{PhysRevLett.86.3200}
\myfigurelarge{\includegraphics[width=0.59\textwidth,angle=-90]{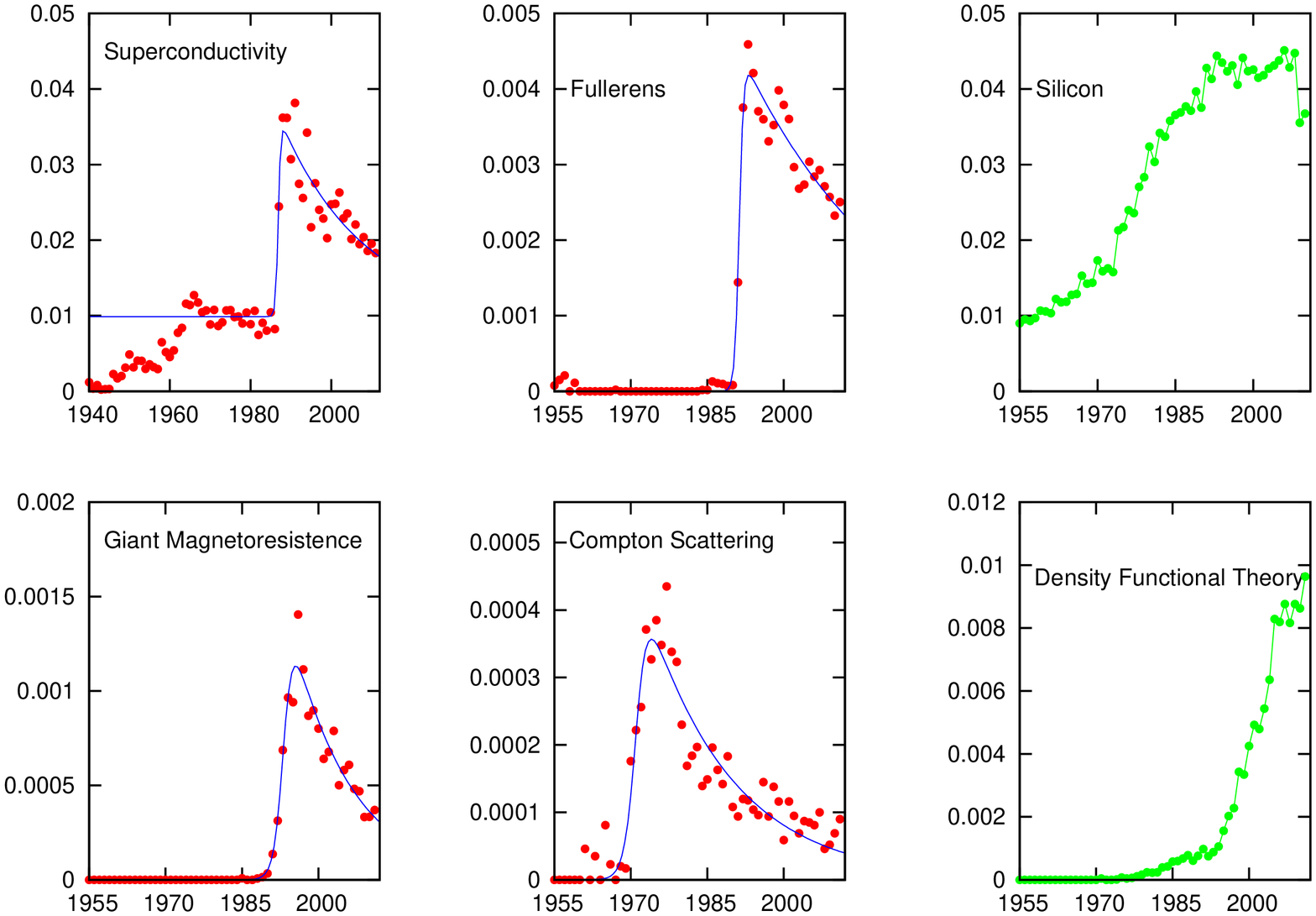}
\figcaption{(color online) Number of papers for different research topics versus the years normalized to the total one in physics. The blue solid lines represent the fitting curves, fitting parameters are reported in Table~\ref{table1}. In green two topics that do not fit in our model.}\label{example4}}

Then we analyze the $\beta$ parameter. In Figure \ref{betayear}  we report the parameter $\beta$ as a function of the maximum position. The maximum position is given by $t_{max} = t_0 + \frac{1}{\alpha+\beta} log (\frac{\alpha }{\beta})$, however since $\alpha$ is usually one or two order of magnitude larger than $\beta$, the maximum position can be in many cases approximated with $t_0$. 
\\ 
The parameter $\beta$ is related to the life-time of a research topic. We found that there is a variation of this parameter before and after the 1995. In fact we have an average life-time of about 25 years before 1995  while the average life-time drops to just 10 years in the last period.
\section{Conclusions and open questions}
\label{conclusion}
In conclusion we present a simple model to describe trends in science and applied it to condensed matter physics. We found that different research topics are well represented by our model. In particular topics that had limited applications in research or industry. These topics show a fast growth followed by an exponential decay in publication ratio. On the contrary other topics that became relevant for industry (as Silicon, transistors, lasers, etc..) and/or for basic research (as lasers, many-body perturbation theory, density functional theory,...) show a growth comparable or superior to the total publications number in physics. We learn from these results that while new topics spread  in the scientific community in few years their decay time is of the orders of decades.\\
\myfigure{\includegraphics[width=0.7\textwidth,angle=-90]{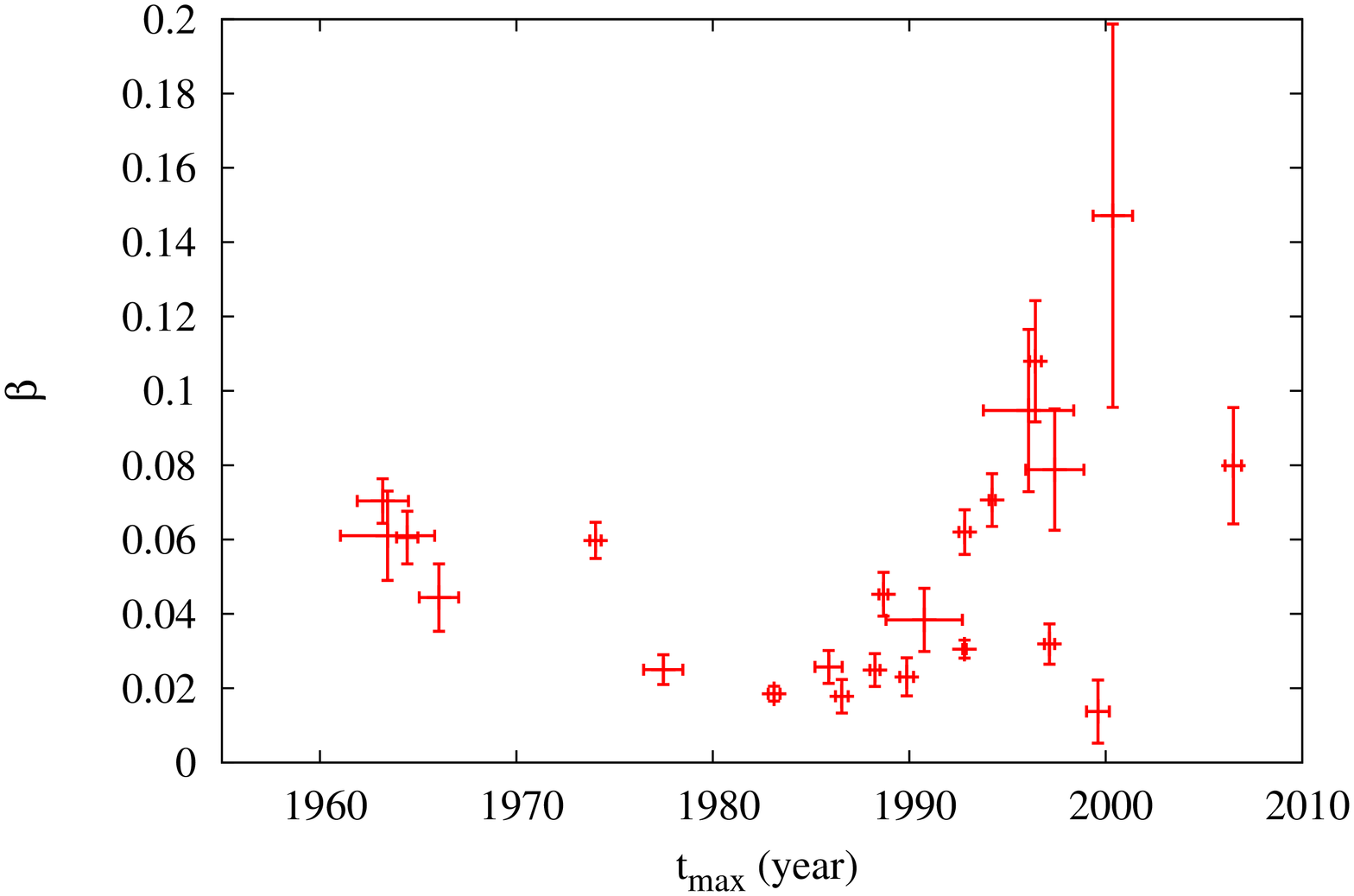}
    \figcaption{(color online) The $\beta$ parameter as a function of $t_{max}$ that represents the peak maximum position.}
\label{betayear}}
These results could be used for example to investigate "scientific bubbles"\cite{Loeb1} or to discriminate between the importance of different sub-fields\cite{Loeb2}. In fact one can expect that "bubbles" will have a different decay in time than real research sub-fields, and try to detect them. Then we analyzed the $\alpha$ and $\beta$ parameters that are related to the birth and death of a research sub-field. We found an increase of both $\beta$ and  $\alpha$ with time. Unfortunately the data are too noisy to model this behavior and to draw reliable conclusions. Therefore the question we posed in the title remains open: is research going faster and faster? Or in another way are trends in condensed matter physics changing faster than in the past? We have some indications but not a definite answer. \\
For sure there are different possibilities to improve the present results that we have not considered yet, as for instance to extend this research to all physics branches, including more sources or combining lexical analysis with citation metrics.                                  We hope this paper will stimulate future research on diffusion, birth and depth of trends in the scientists network.
\section*{Acknowledgments}
CA acknowledges Lorenzo Stella for useful discussions, and Elena Cannuccia for carefully reading this paper. This work has not been supported by any funding program.

\addcontentsline{toc}{chapter}{Bibliography}
\bibliographystyle{unsrt}
\bibliography{FinP}

\begin{thebibliography}{10}

\bibitem{RevModPhys.81.591}
Claudio Castellano, Santo Fortunato, and Vittorio Loreto.
\newblock Statistical physics of social dynamics.
\newblock {\em Rev. Mod. Phys.}, 81:591--646, May 2009.

\bibitem{RevModPhys.74.47}
R\'eka Albert and Albert-L\'aszl\'o Barab\'asi.
\newblock Statistical mechanics of complex networks.
\newblock {\em Rev. Mod. Phys.}, 74:47--97, Jan 2002.

\bibitem{newman}
M.~E.~J Newman.
\newblock The structure of scientific collaboration networks.
\newblock {\em PNAS}, 98(2):404, 2001.

\bibitem{PhysRevLett.86.3200}
Romualdo Pastor-Satorras and Alessandro Vespignani.
\newblock Epidemic spreading in scale-free networks.
\newblock {\em Phys. Rev. Lett.}, 86:3200--3203, Apr 2001.

\bibitem{glanzel2012bibliometric}
W.~Gl{\"a}nzel.
\newblock Bibliometric methods for detecting and analysing emerging research
  topics.
\newblock {\em El profesional de la informaci{\'o}n}, 21(2):194--201, 2012.

\bibitem{springerlink:10.1007/BF02017232}
M.~Zitt and E.~Bassecoulard.
\newblock Development of a method for detection and trend analysis of research
  fronts built by lexical or cocitation analysis.
\newblock {\em Scientometrics}, 30:333--351, 1994.

\bibitem{weidlich}
Wolfgan Weidlich.
\newblock Physics and social science: the approach of synergetics.
\newblock {\em Phys. Rep.}, 204:1--163, 1991.

\bibitem{Nakayama}
Shoichiro Nakayama and Yasuyuki Nakamura.
\newblock A fashion model with social interaction.
\newblock {\em Physica A: Statistical Mechanics and its Applications}, 337:625,
  2004.

\bibitem{expgrowth}
Jean Tague, Jamshid Beheshti, and Lorna Rees-Potter.
\newblock The law of exponential growth: Evidence, implications and forecasts.
\newblock {\em Library Trends}, 30:125, 1981.

\bibitem{SAO}
ADS, CfA, and NASA.
\newblock Sao/nasa astrophysics data system abstract service.

\bibitem{data}
All data are obtained from the SAO/NASA abstract service\cite{SAO} by mean of a
  Python script that automatically performs the researchs and parses the
  results. We tried also to perform research on other databases as ISI Web of
  Science, APS journals or arXiv and we found a similar behaviour. However data
  in SAO/NASA are larger than the other mentioned databases and this allows to
  reduce fluctuation among the different years.

\bibitem{0034-4885-32-2-306}
L~J Anthony, H~East, and M~J Slater.
\newblock The growth of the literature of physics.
\newblock {\em Reports on Progress in Physics}, 32(2):709, 1969.

\bibitem{logistic}
Pierre-Fran\c{c}ois Verhulst.
\newblock Notice sur la loi que la population poursuit dans son accroissement.
\newblock {\em Correspondance math\'ematique et physique}, 10:113, 1838.

\bibitem{economyscience}
Stephan Paula~E.
\newblock The economics of science.
\newblock {\em Journal of Economic Literature}, 34(3):1199, 1996.

\bibitem{Yukalov20091752}
V.I. Yukalov, E.P. Yukalova, and D.~Sornette.
\newblock Punctuated evolution due to delayed carrying capacity.
\newblock {\em Physica D: Nonlinear Phenomena}, 238(17):1752 -- 1767, 2009.

\bibitem{Loeb1}
Abraham Loeb.
\newblock Rating growth of scientific knowledge and risk from theory bubbles.
\newblock http://arxiv.org/abs/1108.5282, 2011.

\bibitem{Loeb2}
Abraham Loeb.
\newblock How to nurture scientific discoveries despite their unpredictable
  nature.
\newblock http://arxiv.org/abs/1207.3812, 2012.

\end{thebibliography}

\end{multicols}


\end{document}